\documentclass[%
 reprint,
 amsmath,amssymb,
 aps,
]{revtex4-2}

\usepackage{svg}
\usepackage{comment}
\usepackage{upgreek}
\usepackage{graphicx}
\usepackage{dcolumn}
\usepackage{bm}
\begin{document}

\preprint{APS/123-QED}

\title{Photonic indistinguishability of the tin-vacancy center in nanostructured diamond}

\author{Jes\'{u}s Arjona Mart\'{i}nez$^{1,*}$}
\author{Ryan A. Parker$^{1,*}$}
\author{Kevin C. Chen$^{2}$}
\author{Carola M. Purser$^{1}$}
\author{Linsen Li$^{2}$}
\author{Cathryn P. Michaels$^{1}$}
\author{Alexander M. Stramma$^{1}$}
\author{Romain Debroux$^{1}$}
\author{Isaac B. Harris$^{2}$}
\author{Martin Hayhurst Appel$^{1}$}
\author{Eleanor C. Nichols$^{1}$}
\author{Matthew E. Trusheim$^{2}$}
\author{Dorian A. Gangloff$^{1,3,\dagger}$}
\author{Dirk Englund$^{2,\dagger}$}
\author{Mete Atatüre$^{1,\dagger}$}
\affiliation{%
 ${}^{1}$Cavendish Laboratory, University of Cambridge, JJ Thomson Avenue, Cambridge CB3 0HE, United Kingdom
}%
\affiliation{%
 ${}^{2}$Department of Electrical Engineering and Computer Science, Massachusetts Institute of Technology, Cambridge, Massachusetts 02139, USA
}%
\affiliation{%
 ${}^{3}$Department of Engineering Science, University of Oxford, Parks Road, Oxford, OX1 3PJ
} 

\date{\today}

\begin{abstract}

Tin-vacancy centers in diamond are promising spin-photon interfaces owing to their high quantum-efficiency, large Debye-Waller factor, and compatibility with photonic nanostructuring. Benchmarking their single-photon indistinguishability is a key challenge for future applications. Here, we report the generation of single photons with $99.7^{+0.3}_{-2.5}\%$ purity and $63(9)\%$ indistinguishability from a resonantly excited tin-vacancy center in a single-mode waveguide. We obtain quantum control of the optical transition with $1.71(1)$-ns-long $\pi$-pulses of $77.1(8)\%$ fidelity. A modest Purcell enhancement factor of 12 would enhance the indistinguishability to $95$\%. The greater than $100$ ms spectral stability shown would then enable strings of up to $10^6$ identical photons to be generated.

\end{abstract}

\maketitle


Indistinguishable photons from quantum emitters provide a fundamental resource for scalable quantum communication and have been employed to realize linear optical quantum computation \cite{barrett2005linearopticalcomputing, lanyon2013measurement, wang2019bosonsampling, barrett2005linearopticalcomputing}, spin-photon and spin-spin entanglement \cite{cabrillo1999cabrillociracscheme, vasconcelos2020nvtimetopolarization, bernien2013nvremoteentanglement}, and quantum repeater schemes \cite{yuan2008experimental, azuma2015photonicquantumrepeater, li2019experimental, lago2021telecomrepeater}. Experimentally, the photonic indistinguishability can be benchmarked through two-photon quantum interference known as the Hong-Ou-Mandel (HOM) effect. The HOM indistinguishability places a bound on the fidelities achievable in photon-mediated gates and entanglement distribution in measurement-based protocols \cite{hong1987hongoumandel, zwerger2016measurementquantumcommunication, rudolph2017optimistic}. This effect has been observed across multiple solid-state emitters such as the nitrogen-, silicon- and germanium-vacancies in diamond \cite{bernien2012homnv, sipahigil2014sivhom, chen2022gevhom}, defects in silicon-carbide \cite{morioka2020spin}, and semiconductor quantum dots \cite{santori2002homqd}.

Within solid-state emitters, the negatively charged group-IV centers in diamond stand as promising spin-photon interfaces due to their large Debye-Waller factor (60\% - 80\%) \cite{Neu2011FluorescenceIridium, Palyanov2015, bhaskar2017quantum, gorlitz2020spectroscopy}, competitive quantum efficiency (10\% - 80\%) \cite{sukachev2017siv, gorlitz2020spectroscopy, bhaskar2017quantum, iwasaki2017tindiscovery} and first-order insensitivity to electric-field noise \cite{desantis2021snvstark, thiering2018abinitiodft}. This electric-field agnosticism makes them inherently compatible with complex photonic nanostructuring, as the disordered charge environment present at the diamond interface does not couple deleteriously to a proximate emitter. Accordingly, high collection efficiency \cite{knall2022sivhighcollection, bhaskar2020sivcavity}, Purcell enhanced emission \cite{rugar2021snvcavity, kuruma2021snvcavity}, and incorporation into photonic integrated circuits \cite{wan2020photoniccircuits} have been demonstrated using group-IV centers embedded in diamond. The negatively charged tin-vacancy (SnV) is particularly promising. Its large ground-state orbital splitting inhibits phonon-mediated dephasing \cite{bradac2019aharonovichreview}, allowing for operation at a temperature accessible in standard helium closed-cycle cryostats. Accordingly, a spin-coherence time ($T_2$) of $0.30(8) \; \text{ms}$ at $1.7 \; \text{K}$ has been achieved using a SnV center \cite{debroux2021quantumcontrol}; longer than that of other group-IV emitters at this temperature. Observation of transform-limited emission in bulk diamond \cite{trusheim2020transformlimited} and advancements in fabrication and charge stability \cite{gorlitz2022chargestability} further demonstrate that SnV centers are a suitable spin-photon interface for quantum networking and measurement-based computation protocols \cite{zwerger2016measurementquantumcommunication}. Probing the optical coherence of the emitted photons is the only major unexplored benchmark for this spin-photon interface.

\begin{figure}
    \centering
    \includegraphics[width=1.0\columnwidth]{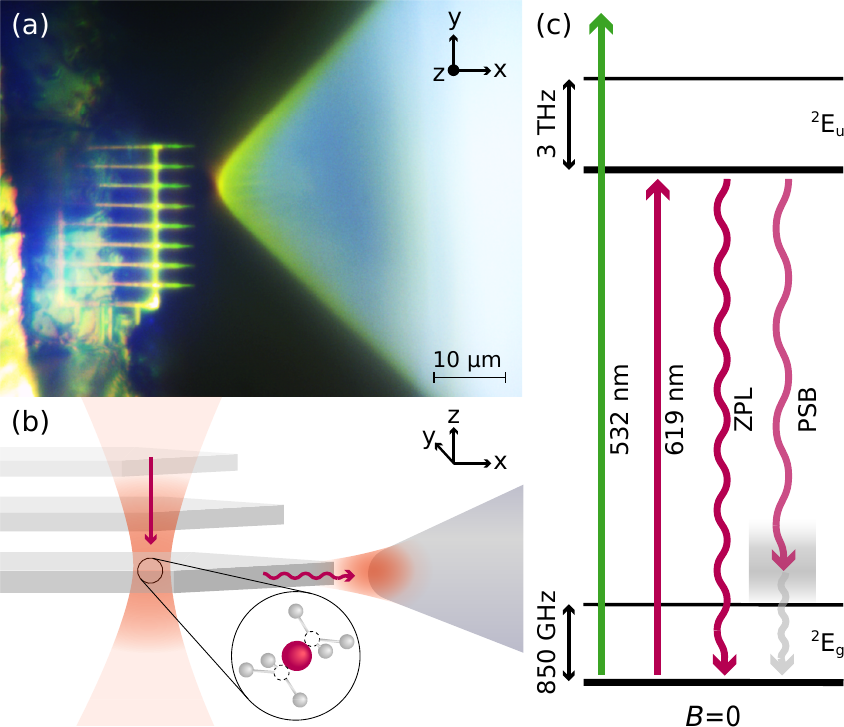}
    \caption{
        (a) Microscope photograph of the waveguide to lensed fiber alignment. The waveguide array is placed on the edge of a silicon substrate such that it protrudes over the edge. A 3-axis nanopositioning stack is used to align a lensed fiber to a single waveguide. (b) Diagrammatic representation of the excitation and collection scheme. The SnV center is excited orthogonally to the waveguide axis but fluorescence is collected through the lensed fiber. (c) Electronic structure of the tin-vacancy center with no magnetic field. Off-resonant 532-nm and on-resonance 619-nm lasers are employed throughout this work to generate emission into the ZPL and PSB radiative decay pathways.
    }
    \label{fig:1}
\end{figure}

In this work, we report the observation of quantum interference of single photons from a SnV center in a single-mode diamond waveguide with an indistinguishability of $63(9)\%$ and a single-photon purity of $99.7^{+0.3}_{-2.5}\%$. This is achieved by realizing coherent control of the optical transition of the SnV center with a $\pi$-rotation fidelity of $77.1(8)\%$ realized in $1.71(1)$ ns. Thus, we show that the SnV center has sufficient photonic coherence to satisfy the requirements for quantum networking.


We perform our experiments on a diamond grown via chemical vapor deposition that has been implanted with Sn$^{++}$ ions (350 keV; 7$^o$ implantation angle; $10^{9}$ cm$^{-2}$ fluence), annealed, and subsequently fabricated into a diamond waveguide chiplet \cite{mouradian2017nanocavityfabrication,hom2022si}. A chiplet consists of eight single-mode waveguides, each $50$~$\upmu$m long with a rectangular cross-section of $200$~nm by $270$~nm. Figure~\ref{fig:1}(a) shows a characteristic device. The waveguides have an adiabatic taper over a distance of $9~\upmu$m at their two ends. This geometry is chosen to maximize the SnV center's emission out-coupling efficiency in finite-difference-time-domain simulations \cite{hom2022si}. A support diamond structure connects the waveguides and provides sufficient structural integrity to allow the chiplet to be pick-and-placed onto the edge of a silicon substrate \cite{wan2020photoniccircuits}. Such placement allows the light emitted by the SnV center to be collected by a single-mode lensed fiber giving a $23(3)$-fold enhancement in collection efficiency relative to conventional confocal microscopy for the emitter studied in this work \cite{hom2022si}.

As indicated in Fig.~\ref{fig:1}(b), the excitation and collection modes are decoupled through orthogonal propagation directions. In this geometry, we achieve a continuous-wave laser suppression in excess of 60 dB. The SnV fluorescence collected through the lensed fiber is then routed from inside the cryostat, held at a temperature of 3.6 K, to our optical setup \cite{hom2022si}.

In the absence of magnetic field, the SnV center has a spin-degenerate optical transition between the lower orbital branch of the ground state and the lower orbital branch of the excited state \cite{hepp2014electronicstructure}. To address this transition resonantly, we employ a $619$-$\text{nm}$ laser, as highlighted in Fig. \ref{fig:1}(c). Also highlighted are the two radiative-decay pathways, the zero-phonon line (ZPL) and phonon sideband (PSB). This resonant drive induces occasional blinking of the SnV center. To remedy this we alternate between resonant and off-resonant $532$-$\text{nm}$ excitation \cite{hom2022si}, which pumps the emitter into the required, photo-active, -1 charge state  \cite{gorlitz2022chargestability}. The studied emitter has an excited-state lifetime $T_1$ of $7.44(20)$ ns, as measured through pulsed excitation, which corresponds to a transform-limited linewidth $\Gamma_0/2 \pi$ of $21.4(2)$ MHz \cite{hom2022si}. This is longer than the $4.5 \; \text{ns}$ reported for bulk diamond \cite{trusheim2020transformlimited}, likely due to the emitter being positioned close to the diamond-air interface \cite{jalas2018emission, gorlitz2020spectroscopy}.

\begin{figure}
    \centering
    \includegraphics[width=1.0\columnwidth]{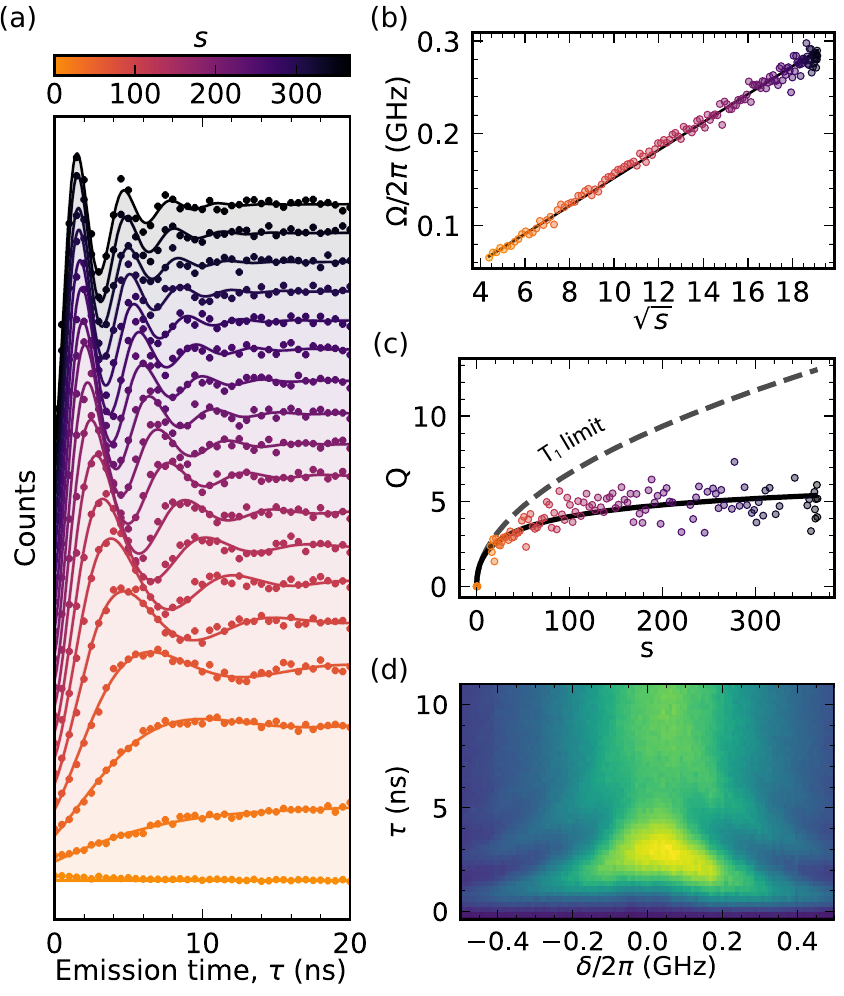}
    \caption{
        (a) Histogrammed PSB fluorescence during a 20-ns-long resonant laser pulse. Curves are shown in a linear scale as a function of the saturation parameter, $s$. Solid curves correspond to individual fits to a master equation described in the main text \cite{hom2022si}. At the highest saturation parameter the excited-state population saturates to $0.5$ at long times. (b) Optical Rabi frequency $\Omega$ extracted from panel a as a function of $\sqrt{s}$. The black line is a linear fit with zero intercept. (c) Quality factor $\text{Q}$ of the Rabi oscillation as a function of $s$. The solid black curve is a fit to a master-equation including a pure dephasing rate proportional to $\Omega$. The dashed black curve shows the absolute coherence limit. (d) Detuning ($\delta$) dependence of the Rabi oscillations at $s=102(4)$ for the same excitation scheme as panel a. PSB fluorescence is color-coded with blue (yellow) corresponding to low (high) fluorescence.
    }
    \label{fig:2}
\end{figure}

We first demonstrate optical control of the SnV center through resonant excitation and collection of the PSB. Figure \ref{fig:2}(a) shows the averaged PSB fluorescence during a fixed-length 20-ns-long resonant 619-nm pulse starting at time $\tau=0$. The histogrammed time-resolved fluorescence throughout the pulse is proportional to the instantaneous excited-state population and shows Rabi oscillations between the ground and excited state levels. The laser power $P$ is parameterized by the saturation parameter $s=P/P_{\text{sat}}$, where $P_\text{sat} = 31(2) \; \text{nW}$ is the resonant laser power at which the Rabi rate $\Omega$ equals $\Gamma_0 / \sqrt{2}$. For each power the excited-state population is fit to a two-level master equation, from which we extract the corresponding $\Omega$ and dephasing rate \cite{hom2022si,gorini1976completely, lindblad1976generators}. At the highest driving power ($s = 367(1)$), it takes $1.71(1) \; \text{ns}$ to perform an optical $\pi$-rotation, much faster than the lifetime ($T_1$) of the emitter, with an extracted fidelity of $77.1(8)\%$.

To investigate the effects of decoherence, the master equation used to fit Fig. \ref{fig:2}(a) includes spontaneous emission and pure dephasing, as well as a shot-to-shot detuning fluctuation resulting in inhomogeneous dephasing~\cite{hom2022si}. 
We confirm in Fig. \ref{fig:2}(b) a direct proportionality of the Rabi rate on the square root of the saturation parameter through the relationship $\Omega = \Gamma_0 \sqrt{s/2}$. This linear relationship extends to our highest probed $s$ and indicates that control-limiting imperfections such as phonon-coupling or multi-level driving do not cause an appreciable deviation from the master-equation model \cite{hom2022si}.
Figure \ref{fig:2}(c) shows the power dependence of the quality factor $Q$, defined as the product of $\Omega$ and the $1/e$ envelope decay time extracted from the master-equation fits. In the low excitation power regime, the quality factor increases with power and agrees with the limit set by the finite $T_1$ lifetime. At higher powers the quality factor saturates which implies a laser-induced dephasing mechanism best modeled with a rate that linearly depends on $\Omega$ \cite{bodey2019opticallocking}.

By varying the frequency of the resonant laser, we also probe the dependence of the fluorescence on the detuning $\delta = \omega_\text{l} - \omega_0$, where $\omega_\text{l}$ and $\omega_0$ are the angular frequencies of the 619-nm resonant laser and the transition, respectively. Figure \ref{fig:2}(d) shows the time evolution of the fluorescence as a function of $\delta$ at $s=102(4)$, yielding a Rabi rate of $153(3) \; \text{MHz}$ at $\delta=0$ and faster, lower-amplitude, oscillations for $\delta \neq 0$. The ability to vary the detuning, phase and amplitude of the resonant laser pulses enables multi-axis control of the optical qubit \cite{bodey2019opticallocking,hom2022si}.


\begin{figure}
    \centering
    \includegraphics[width=1.0\columnwidth]{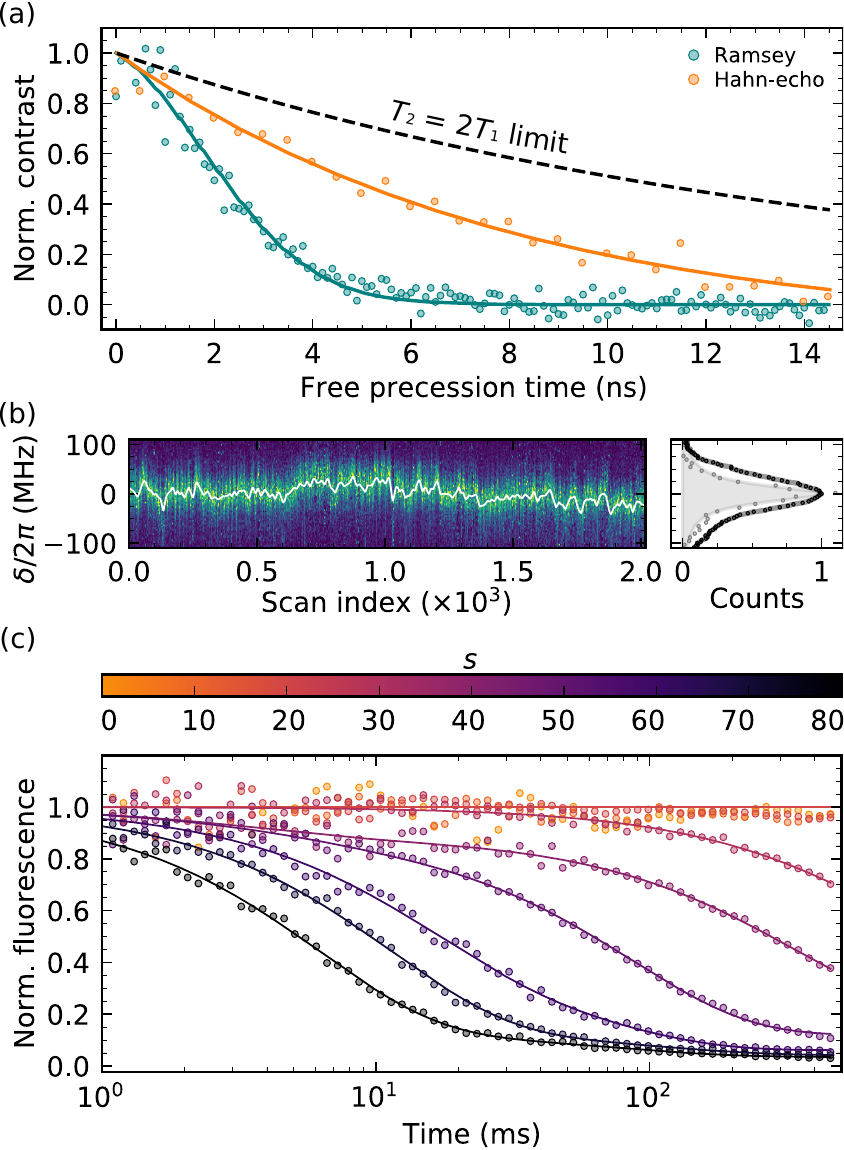}
    \caption{
        (a) Ramsey (teal) and Hahn-echo (orange) contrast decay envelope measurements with master equation fits (solid curves) \cite{hom2022si}. The dashed black curve shows the absolute coherence limit of $T_2 = 2 T_1$. For visual clarity, the contrast at zero free-precession time is normalized to one.
        (b) Left: Shot-to-shot evolution of the PLE lineshape. Each vertical cut shows a histogram resulting from fast repeated PLE scans averaged over 500 ms. Off-resonant 532-nm excitation is applied between resonant sections. The solid white line follows the emitter's resonance frequency. Right: An example line-cut (gray) and overall inhomogeneous distribution of all line-cuts (black), with Lorentzian and Gaussian fits respectively.
        (c) Average PLE intensity as a function of time following a 532-nm off-resonant reset pulse for multiple saturation parameters. Solid curves are fits to bi-exponential decays.
    }
    \label{fig:3}
\end{figure}

We next leverage multi-axis control to probe the coherence of the optical transition directly through pulsed resonant excitation. Figure \ref{fig:3}(a) displays the measurements of Ramsey interferometry and Hahn-echo dynamical decoupling. We read out the state of the emitter by integrating the fluorescence after the final $\pi/2$-rotation. Varying the phase of this $\pi/2$-rotation provides the means to measure the population contrast between the ground and excited states \cite{debroux2021quantumcontrol, bodey2019opticallocking}. The solid teal curve in Fig.~\ref{fig:3}(a) is a fit to the Ramsey-interferometry data using the master-equation model employed previously \cite{hom2022si}. This yields an inhomogeneous dephasing time $T_2^*$ of $4.54(2) \; \text{ns}$, corresponding to a shot-to-shot spectral drift of $82.6(5) \; \text{MHz}$ at full width at half maximum (FWHM). The Hahn-echo contrast decay envelope provides a measurement of the pure-dephasing rate $\Gamma_\text{PD}$. A fit to the raw contrast results in $\Gamma_\text{PD}/2\pi = 6.39(14) \; \text{MHz}$ and an inferred homogeneous linewidth of $\Gamma/2\pi = (\Gamma_0 + 2\Gamma_\text{PD})/2\pi = 34.8(7)$~$\text{MHz}$. This linewidth is only a factor of $1.63(4)$ from its intrinsic Fourier limit ($\Gamma_0$) despite the emitter's close proximity ($<60$~nm) to nanostructured surfaces.

\begin{figure*}
    \centering
    \includegraphics[width=1.0\textwidth]{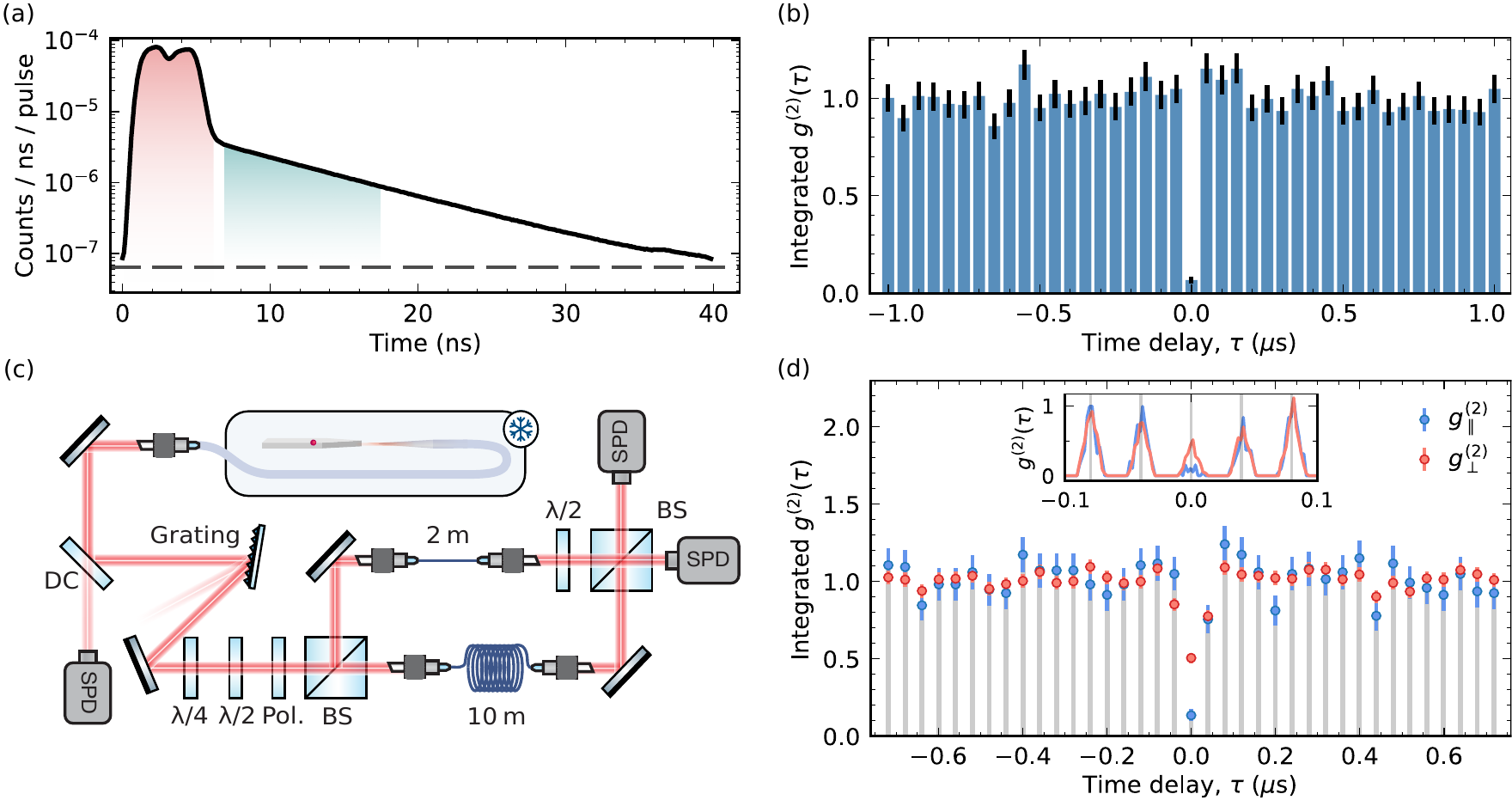}
    \caption{
        (a) Time-resolved ZPL fluorescence under excitation with a 4-ns-long resonant laser pulse. Red and teal shadings correspond to the regions when the laser excitation is present and when the ZPL fluorescence is collected, respectively. The dashed line corresponds to the average background, as extracted from a far detuned ($\delta/2\pi = 6~\text{GHz}$) laser pulse \cite{hom2022si}. (b) Second-order autocorrelation of the ZPL fluorescence under pulsed resonant excitation. The recorded value $g^{(2)}(0)_\text{raw} = 0.067(23)$ corresponds to a background-corrected photon purity of $99.7^{+0.3}_{-2.5}\%$. (c) Experimental setup for the HOM measurement. Abbreviations: quarter-wave plate ($\lambda$/4), half-wave plate ($\lambda$/2), polarizer (Pol.), beamsplitter (BS) and dichroic mirror (DC). (d) Pulsed two-photon interference measurement for the parallel (blue) and perpendicular (red) polarization configurations, where the coincidence counts are integrated for each pulse. We extract a background-corrected photon indistinguishability of $V = 73(13)\%$ for photons collected within the teal region of panel a. Inset: time-resolved distribution of coincidences around $\tau=0$.
    }
    \label{fig:4}
\end{figure*}

We next probe the spectral stability on longer timescales by observing the shot-to-shot variation of the transition frequency through fast photoluminescence excitation (PLE) scans at $s=0.8(1)$. Each scan alternates between $500 \; \text{ms}$ of resonant $619$-$ \text{nm}$ and off-resonant $532$-$\text{nm}$ excitation. During the resonant section, $\delta / 2\pi$ is repeatedly scanned from $-100$~MHz to $100$~MHz using $3$-$\upmu\text{s}$-long linearly-chirped laser pulses. Figure \ref{fig:3}(b) shows the spectral evolution over 34 minutes. Within each vertical cut, and after correcting for power broadening, the linewidth is  $35(10)~\text{MHz}$ which is commensurate with the homogeneous linewidth inferred from the Hahn-echo measurement. This indicates that there are no significant, additional, dephasing mechanisms present between the tens of nanoseconds timescale probed in Fig.~\ref{fig:3}(a), and the $500$-$\text{ms}$ timescale probed by the PLE scans in Fig. \ref{fig:3}(b).

Over multiple cuts, the central frequency, extracted from Lorentzian fits, does not show significant variation over consecutive scans. Over the entire measurement, the central frequency variation tends towards a normal distribution with a FWHM of $64(1) \; \text{MHz}$.  Such inhomogeneous broadening has been observed for the nitrogen-vacancy center \cite{bernien2012homnv}, and more recently the SnV center \cite{gorlitz2022chargestability}, and is likely due to the off-resonant laser rearranging the local charge environment. 
For the nitrogen-vacancy center, the photo-induced spectral broadening can be over $0.5$~GHz, or $35$-times broader than the intrinsic linewidth \cite{siyushev2013optically}. In contrast, electric noise only detunes the SnV center through a second-order Stark shift \cite{desantis2021snvstark} leading to a significantly narrower inhomogeneous linewidth. Figure~\ref{fig:3}(b) shows an inhomogeneous broadening of a factor of $\sim 2$ compared to its homogeneous linewidth.

The $532$-nm off-resonant excitation serves to photo-stabilize the emitter as resonant excitation pumps the emitter stochastically into a dark state \cite{gorlitz2022chargestability}. To establish the timescale of this pumping, we monitor the decay of PLE intensity following a $532$-nm off-resonant pulse. Repeatedly scanning $300~\text{MHz}$ across the transition makes the measurement insensitive to spectral diffusion. Figure~\ref{fig:3}(c) shows the decay of PSB fluorescence for various resonant excitation powers. The solid curves in Fig.~\ref{fig:3}(c) are fits to a bi-exponential model \cite{parker2021infrared}, where the dominant component reveals the time scale of pumping into the dark state. At the highest laser power, fluorescence persists over milliseconds allowing an average of $\sim 10^{6}$ $\pi$-pulses to be applied before pumping into the dark state. When considered in concert with Fig. \ref{fig:3}(b), our emitter remains spectrally stable and optically active on a many millisecond timescale as required for quantum networking \cite{pompili2021nvmultinodenetwork}. 


To probe the indistinguishability of the emitted photons, we isolate the ZPL using spectral filtering. We employ two electro-optic modulators for fast optical pulsing of the resonant excitation laser and time-tag the ZPL fluorescence. This reduces the background from laser scattering in excess of $30$~dB, which combined with the orthogonal excitation and collection directions results in a total laser suppression greater than $90$~dB. Figure \ref{fig:4}(a) shows the time-resolved ZPL fluorescence after a $4$-ns-long resonant excitation pulse. The large laser suppression yields a signal-to-background ratio of $23.91(3)$ during the first $11.1~\text{ns}$ following the excitation pulse, which corresponds to $80(1)\%$ of the integrated ZPL fluorescence \cite{hom2022si}.

Routing the ZPL emission through a Hanbury-Brown and Twiss interferometer \cite{brown1956hanburybrowntwiss}, we measure a zero-delay second-order intensity correlation $g^{(2)}(0)_
\text{raw}$ of $0.067(23)$, as shown in Fig. \ref{fig:4}(b). This is comparable to previous reports where only the PSB was collected, despite the added technical challenge of resonant collection in this work \cite{trusheim2020transformlimited, debroux2021quantumcontrol}. Correcting for our detector dark-count rates, we calculate a background-corrected photon purity ($1-g^{(2)}(0)$) of $99.7^{+0.3}_{-2.5}\%$.

Figure \ref{fig:4}(c) shows our HOM-interferometry setup \cite{hom2022si}. Before the interferometer, we employ a dichroic mirror and a grating to filter photons within a 245(6) GHz bandwidth (FWHM) centered on the ZPL transition. This filtered emission is routed through polarization control optics and into the interferometer with a relative time delay $\Delta t$ of $39.93(2)~\text{ns}$ between the two arms. To ensure temporal overlap between subsequently emitted photons we apply laser pulses with a repetition period matching $\Delta t$. A half-wave plate placed in the short-arm of the interferometer controls the relative polarization between the two interfering photons.

Figure \ref{fig:4}(d) shows the second-order intensity correlation measured across the two output ports of the interferometer for ZPL fluorescence integrated over the first $11.1$~ns after the excitation pulse. When the polarizations of two interferring photons are orthogonal to each other, we measure $g_\perp^{(2)}(0) = 0.51(3)$, in agreement with the theoretically expected value of 0.5. When the polarizations of the photons are matched, we measure $g_\parallel^{(2)}(0) = 0.22(3)$. The raw visibility, $V_\text{raw} = 1 - g_\parallel^{(2)}(0)/g_\perp^{(2)}(0) = 56(8)\%$, is a measurement of the indistinguishability of the emitted photons. Correcting for the background and finite classical interferometric contrast, we calculate a  photon indistinguishability $V$ of $73(13)\%$. When the time window is extended to the full decay observed in Fig~\ref{fig:4}(a) the indistinguishability is reduced to $V = 63(9)\%$ \cite{hom2022si}.

We compare our measured indistinguishability to a model that includes homogeneous broadening as the source of photon distinguishability. Given the measured $\Gamma_{\text{PD}}$, we compute an expected HOM visibility $V_\text{sim}$ of $63.2(4)\%$ in agreement with our measured value \cite{hom2022si}. Considering the slow spectral diffusion measured in Fig~\ref{fig:3}(b), the HOM visibility measured here 
should extend to interferometer delays greater than $100$~ms.


In this work, we demonstrate coherent control of the optical transition of the SnV in diamond. We achieve a $\pi$-pulse time of $1.71(1)~\text{ns}$ and two-photon quantum interference of resonant photons with an indistinguishability of $63(9)\%$. Purcell enhancement of the emitter, such as through optical microcavities \cite{riedel2017nvmicrocavity} or photonic crystal cavities  \cite{bhaskar2020sivcavity,rugar2021snvcavity, kuruma2021snvcavity}, increases the radiative decay rate and thus reduces the sensitivity to dephasing. An overall Purcell enhancement factor of 12, already achieved using SnV centers in nanocavities \cite{rugar2021snvcavity, kuruma2021snvcavity}, would yield a photon indistinguishability in excess of 95\%. Moreover, fabrication and material processing improvements, known to result in improved optical quality of SnV centers in bulk diamond \cite{gorlitz2020spectroscopy}, should enhance the optical coherence. These advancements provide a feasible route to near-unity photon indistinguishability in the near-term. 

Using the two-photon interference presented in this work, in combination with control of the spin degree of freedom \cite{debroux2021quantumcontrol}, one could realize spin-photon entanglement \cite{vasconcelos2020nvtimetopolarization, togan2010nvspinphotonentanglement, gao2012qdspinphotonentanglement} and entanglement of remote emitters \cite{cabrillo1999cabrillociracscheme, humphreys2018nvspinspinentanglement, stockill2017qdspinspinentanglement}. Modest improvements in the indistinguishability of the photons would position multi-photon entangled states as the next achievement using the SnV center \cite{buterakos2017repeatergraphstate, michaels2021multidimensionalclusterstate}.

\begin{acknowledgments}
We acknowledge support from the ERC Advanced Grant PEDESTAL (884745), the EU Quantum Flagship 2D-SIPC. J.A.M. acknowledges support from the Winton Programme and EPSRC DTP, R.A.P. from the General Sir John Monash Foundation, K.C.C. from the National Science Foundation Graduate Research Fellowships Program (GRFP) and the NSF STC Center for Integrated Quantum Materials (CIQM), NSF Grant No. DMR-1231319 and NSF Award No. 1839155, C.P.M. from the EPSRC DTP, A.M.S. from EPSRC/NQIT, R.D. from the Gates Cambridge Trust, M.T. from the Army Research Laboratory ENIAC Distinguished Postdoctoral Fellowship, and D.A.G. from a St John's College Title A Fellowship and a Royal Society University Research Fellowship. D.E. acknowledges further support by the MITRE Quantum Moonshot Program.

${}^{*}$ These authors contributed equally to this work.

${}^{\dagger}$ Correspondence should be addressed to: \href{mailto:dorian.gangloff@eng.ox.ac.uk}{dorian.gangloff@eng.ox.ac.uk}, \href{mailto:englund@mit.edu}{englund@mit.edu}, \href{ma424@cam.ac.uk}{ma424@cam.ac.uk}.

\end{acknowledgments}

\end{document}


\preprint{APS/123-QED}

\title{Supplemental information for \\ 
Photonic indistinguishability of the tin-vacancy center in nanostructured diamond}

\affiliation{\textsuperscript{1}Cavendish Laboratory, University of Cambridge, JJ Thomson Avenue, Cambridge CB3 0HE, United Kingdom}
\affiliation{\textsuperscript{2}Department of Electrical Engineering and Computer Science, Massachusetts Institute of Technology, Cambridge, MA 02139, USA}
\affiliation{\textsuperscript{3}Cambridge Graphene Centre, University of Cambridge, Cambridge CB3 0FA, United Kingdom
\\ \ \\
\textsuperscript{*}\,These authors contributed equally to this work. \\
\textsuperscript{$\dagger$}\,Correspondence should be addressed to: englund@mit.edu, dag50@cam.ac.uk, and ma424@cam.ac.uk. \\}

\maketitle
\tableofcontents

\section{Sample fabrication}

\begin{figure*}[h]
    \centering
    \includegraphics[width=0.5\textwidth]{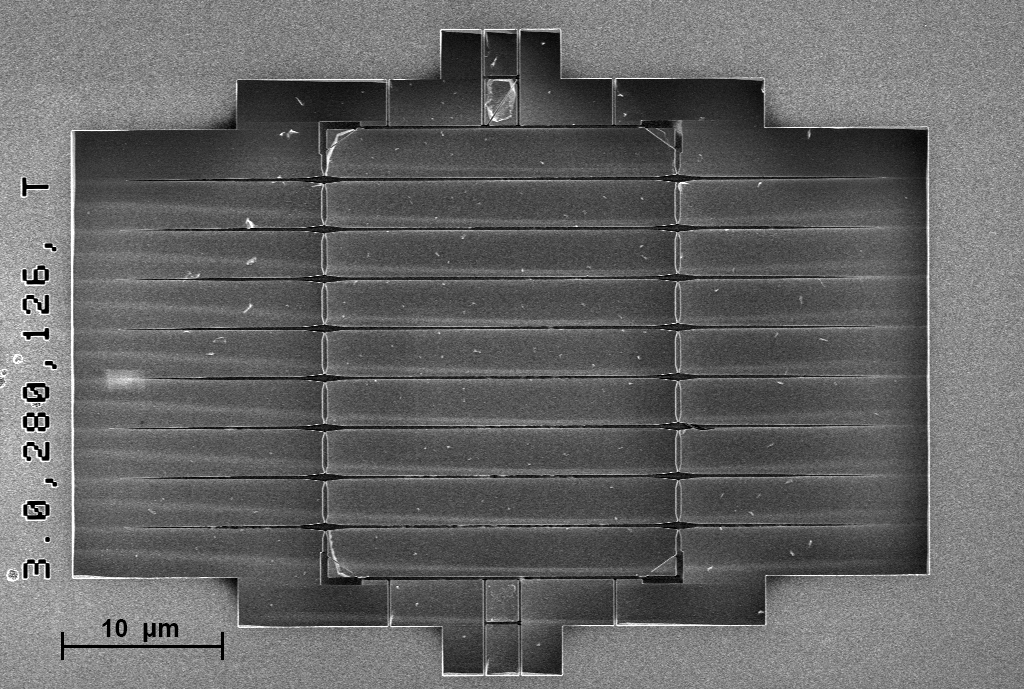}
    \caption{
        Quantum micro-chiplet imaged through Scanning Electron Microscopy (SEM).
    }
    \label{fig:sem}
\end{figure*}

An electronic-grade diamond (Element 6, [N] $< 5$ ppb) was first chemically polished via $\text{ArCl}_2$ Reactive Ion Etching (RIE) to smooth the surface to $<1$ nm roughness, then implanted with $\text{Sn}^{++}$ ions (INNOViON Corp.) at a dosage of $5 \times 10^{11}$ ions/$\text{cm}^2$ and an energy of 350 keV. This corresponds to a nominal $86$ nm implant depth with a $17$ nm straggle, based on SRIM simulations \cite{ziegler1985stopping}. The diamond subsequently underwent high-temperature annealing at 1200 ${}^{\circ}$C for 12 hours at $\sim 10^{-7}$ mbar followed by a boiling tri-acid clean (nitric, sulfuric, and perchloric acids mixed in 1:1:1 ratio and heated to 345 ${}^{\circ}$C). Quantum micro-chiplets (QMC) \cite{wan2020photoniccircuits} each containing 8 waveguide channels were fabricated in bulk diamond through quasi-isotropic etching \cite{mouradian2017nanocavityfabrication, wan2018twodimensionalnanocavity}. A characteristic QMC is shown in figure \ref{fig:sem}. After device fabrication, the sample again underwent high temperature annealing at 1200 ${}^{\circ}$C for 12 hours and subsequent boiling tri-acid clean. In addition, the diamond was submerged in piranha (sulfuric acid mixed with hydrogen peroxide in a 3:1 ratio) to improve surface termination \cite{sangtawesin2019diamondsurfacenoise}. The nanostructures were then transferred onto the edge of a silicon-based substrate via a pick-and-place technique with a fine-tipped tungsten probe \cite{wan2020photoniccircuits, mouradian2015scalablememoryphotonic}. The substrate consisted of 525 $\upmu$m thick silicon, 50 nm thick gold, and 30 nm thick PMMA, which was found to provide sufficient adhesion for the transferred devices to be resilient against mechanical vibrations in transport and thermal cycling.

\section{Characterization of collection efficiency}

\subsection{Simulation}

\begin{figure*}[h]
    \centering
    \includegraphics[width=0.5\textwidth]{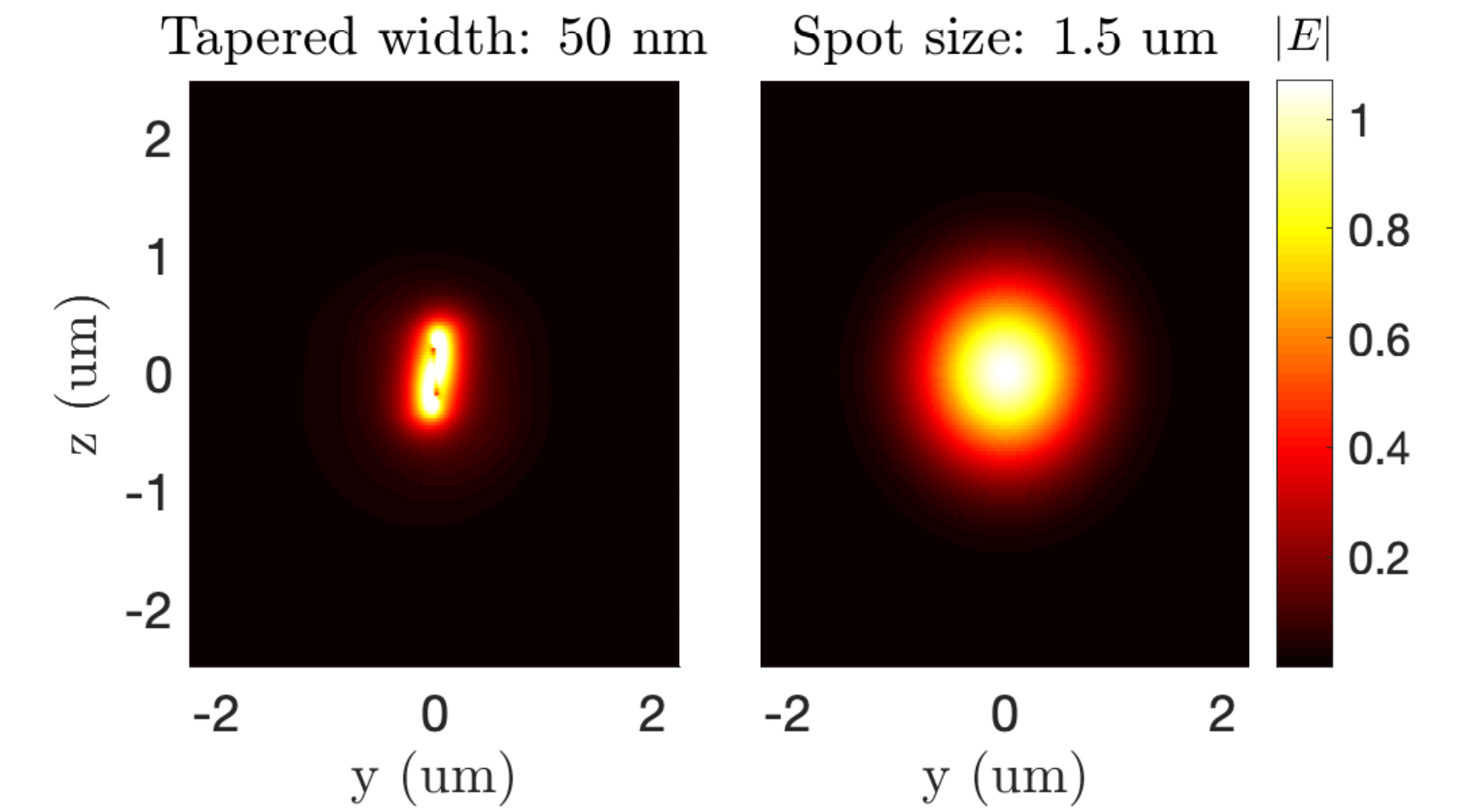}
    \caption{Far-field mode profiles as a function of the two directions orthogonal to the optical axis for (a) a diamond waveguide linearly tapered to a width of 50~nm and (b) a lensed fiber with a spot size of 1.5~$\mu$m.
    }
    \label{fig:mode_sim}
\end{figure*}

Figure \ref{fig:mode_sim}(a) shows the simulated far-field profile of the emission of a SnV coupled to a nanophotonic waveguide of 270~nm width and 200~nm thickness. The mode profile is obtained via finite-difference time-domain (FDTD, Lumerical) simulations. The emitter is modeled as a point dipole aligned to the [111] axis in a [110] top-facing diamond lattice, and located 86~nm in depth. This corresponds to the mean position of the implanted ions. With a waveguide whose width is tapered down linearly to 50~nm over a $\sim 9$~$\mu$m distance, we calculate the coupling efficiency between the waveguide mode and a lensed fiber mode (with a spot size of 1.5~$\mu$m) shown in Fig.~\ref{fig:mode_sim}(b) to be $\sim 77\%$.

\subsection{Experimentally realised collection efficiency}

We experimentally measure our collection efficiency through a pulsed excitation scheme where we collect the PSB fluorescence (SI III). Integrating over the resulting exponential decay allows us to compute a probability of 0.0067(13)\% of collecting a photon in the phonon side band per excitation event. Correcting for the Debye-Waller factor (60\%) \cite{gorlitz2020spectroscopy}, quantum efficiency (80\%) \cite{gorlitz2020spectroscopy, iwasaki2017tindiscovery}, setup efficiency (80(5)\%) and the fact that only half of the photons emitted into the waveguide mode will exit the waveguide through the colllection port we compute a coupling efficiency from the waveguide to the lensed fiber of 0.10(2)\%.

Possible sources that lower the experimentally realised collection efficiency, relative to the simulated efficiency, include a slight angular misalignment between the fiber and waveguide axes, poor axial alignment due to vibrations inside the cryostat, a low coupling factor from the emitter to the waveguide mode ($\beta$) due to the emitter's position, scattering losses from surface roughness, the slightly smaller quantum efficiency of the emitter due to integration into a nanostructure \cite{gorlitz2022chargestability}, or an output mode for the waveguide tapering that deviates significantly from the Gaussian mode of the fiber. 

\subsection{Measured fiber collection enhancement}

\begin{figure*}[h]
    \centering
    \includegraphics[width=1.0\textwidth]{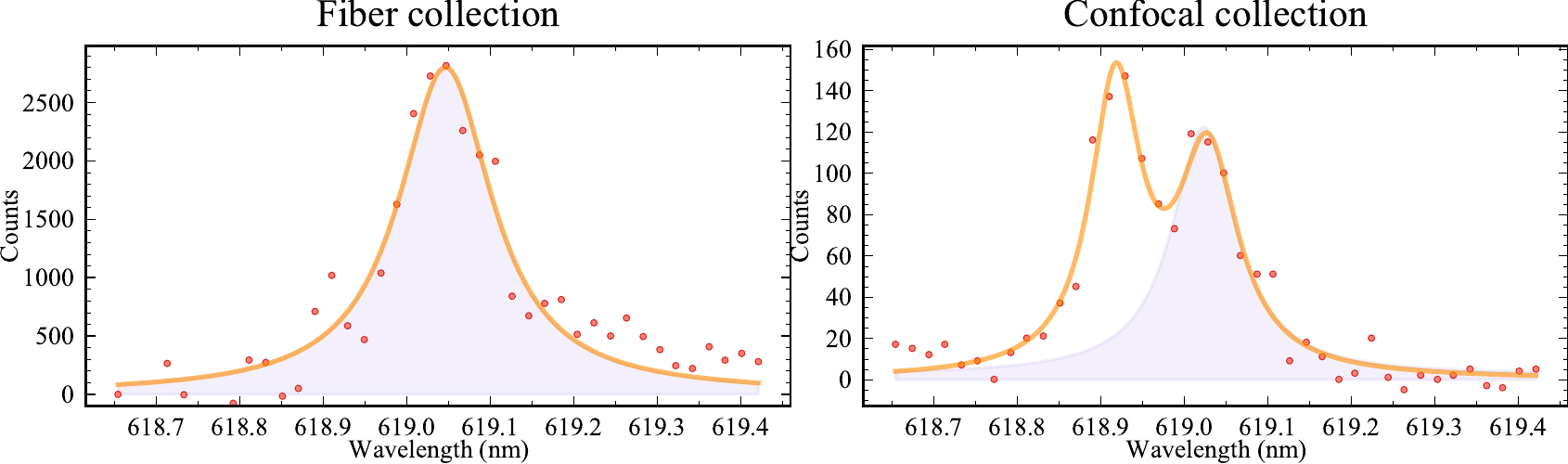}
    \caption{
        Photoluminescence spectra under 532-nm off-resonant illumination collected through the tapered fiber (left) and confocal microscope (right) respectively. The purple shading highlights the emission from the emitter studied in the main text.
    }
    \label{fig:pl}
\end{figure*}

Experimentally we find that, when collecting through the tapered fiber in a geometry shown in Fig. 1(a) of the main text, the collection efficiency is enhanced with respect to confocal collection by a factor of $\times$23(3) for the emitter studied in this work. Figure \ref{fig:pl} shows this enhancement by comparing the photoluminescence intensities when collecting emission through the tapered fiber (left) and through the confocal microscope (right) used to address the emitter. We note that the confocal collection also highlights the presence of another emitter that is not clearly seen when spectra are collected through the fiber. This is a general feature, that the collection enhancement can vary significantly from emitter to emitter, likely owing to the emitter's position within the waveguide, its dipole orientation, and how well its emission couples to the waveguide mode. Irrespective of such variability, we systematically see enhanced collection efficiencies when luminescence is collected through the tapered fiber.

\section{Lifetime measurement}

\begin{figure*}[h]
    \centering
    \includegraphics[width=0.5\textwidth]{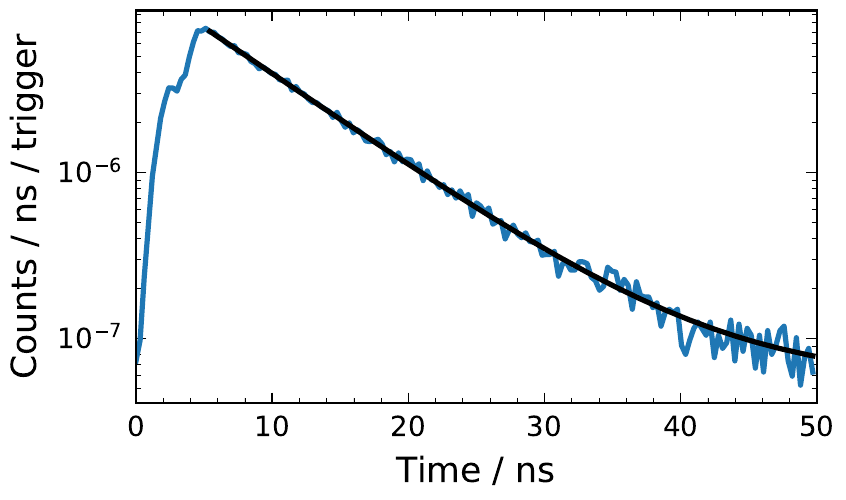}
    \caption{
        Time resolved PSB fluorescence (blue) under excitation with a 4-ns-long resonant laser pulse. The solid black curve corresponds to an exponential fit of the form $A e^{-t/T_1} + B$ to the decay. We extract $A = 1.423(10) \times 10^{-5}$ counts/ns/trigger, $T_1 = 7.44(20)$ ns, and $B=5.6(9) \times 10^{-8}$ counts/ns/trigger.
    }
    \label{fig:lifetime}
\end{figure*}

We perform a measurement of the lifetime of the emitter through pulsed resonant excitation and PSB collection. We apply a 4-ns-long pulse, as in Fig. 4(a) of the main text, but collect the PSB after spectral filtering with a dichroic mirror. From an exponential fit to the spontaneous decay shown in Fig. \ref{fig:lifetime} we extract a lifetime of 7.44(20) ns.

\section{Setup}\label{sec:Setup}

\begin{figure*}[h]
    \centering
    \includegraphics[width=0.8\textwidth]{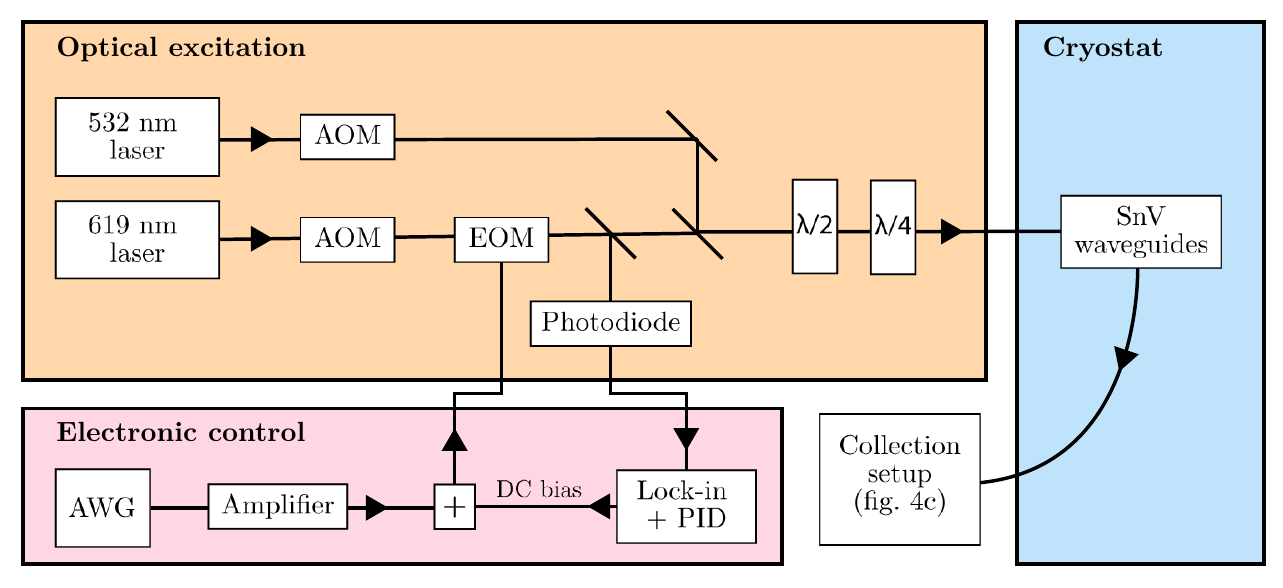}

    \caption{
        Schematic of the experimental setup. 619-nm and 532-nm lasers are pulsed and combined before being directed through an optical port onto the SnV chiplets inside the cryostat. The fluorescence of the emitter is then collected via a fiber and routed to the collection setup shown in Fig~4(c) of the main text. 
    }
    \label{fig:setup}
\end{figure*}

Figure \ref{fig:setup} shows a schematic diagram of our experimental setup. We employ a 619-nm tunable laser (M2 SolsTiS + External Mixing Module) as the coherent resonant light source. This laser light goes through an acousto-optic modulator (AOM, Gooch and Housego 3080-15), used to modify the power transmitted, and a fiber-based electro-optic modulator (EOM, Jenoptik AM635), used to generate fast optical pulses in the carrier (Fig. 4) and in the $\sim 2$ GHz frequency sidebands (Figs. 2-3). To maximise the laser background suppression, the measurement in Fig. 4 in the main text includes a second EOM (not shown) immediately after the first EOM. The electro optic modulators are controlled by a 25 Gsamples/sec arbitrary waveform generator (Tektronix AWG70002A) and locked to their interferometric minimum by a field programmable gate array (Red Pitaya, STEMlab 125-10) employed as a sine wave generator, lock-in amplifier and PID loop.

After the EOM, the light is mixed at a dichroic mirror (Thorlabs DMLP638) with a 532 nm fixed-frequency laser source (M2 Equinox) that can be pulsed with an AOM (Gooch and Housego 3080-15). The pulsing of this AOM as well as triggering of all other equipment is handled by a digital pattern generator (Swabian Instruments Pulse Streamer 8/2). After recombination, the light is routed to the sample via polarization maintaining fibers,  polarization control elements (polarizer, half-wave plate, quarter-wave plate), and a confocal microscope setup with a 70:30 non-polarising cube beamsplitter. The polarization of the resonant excitation beam is matched to the axis of the emitter by minimising the saturation power $P_\text{sat}$.

Inside the cryostat (Bluefors LD250 dilution refrigerator, operating at 3.6 K) we employ an apochromatic cryo-compatible objective (Attocube LT-APO/VISIR/0.82, Numerical Aperture = 0.82). Both the sample and fiber are aligned to each other and to the imaging field by means of a set of 3-axis stick-slip nanopositioners (2x ANPx101/LT and ANPz101/LT). The tapered fiber (Nanonics tapered lensed fiber SM600, working distance = $4 \pm 1~\upmu$m, spot diamater = $0.8^{+0.3}_{-0.0}~\upmu$m) is routed outside the fridge via a vacuum-sealed interface and spliced onto a standard single-mode SM600 fiber (Thorlabs). No magnetic field inside the cryostat is employed at any point in this letter.

The collected SnV emission is separated into the phonon sideband (PSB) and zero-phonon line (ZPL) using a dichroic mirror (DMLP638). In Figs. 2 and 3 of the main text we exclusively analyze the PSB. Here an additional long-pass filter provides further suppression of the resonant light (Semrock BLP01-633R-25, omitted in the main text for clarity) and a single fiber-coupled single-photon-detector (Excelitas SPCM-AQRH-16-FC) provides photon detection. 
For the ZPL detection in Fig. 4 of the main text we make use of the interferometric setup presented in Fig. 4(d) which includes a plane ruled reflection grating (Richardson Gratings 1200 Grooves, 12.5 x 12.5mm, 750nm), polarisation control optics, fiber patch cables (Thorlabs 488PM fiber), and two detectors (Excelitas SPCM-AQRH-16-FC). In Fig. 4(a) and Fig. 4(b), ZPL fluorescence is collected from the two optical paths immediately after the first 50:50 beamsplitter.

\section{Pulse sequences}

\begin{figure*}[h]
    \centering
    \includegraphics[width=0.85\textwidth]{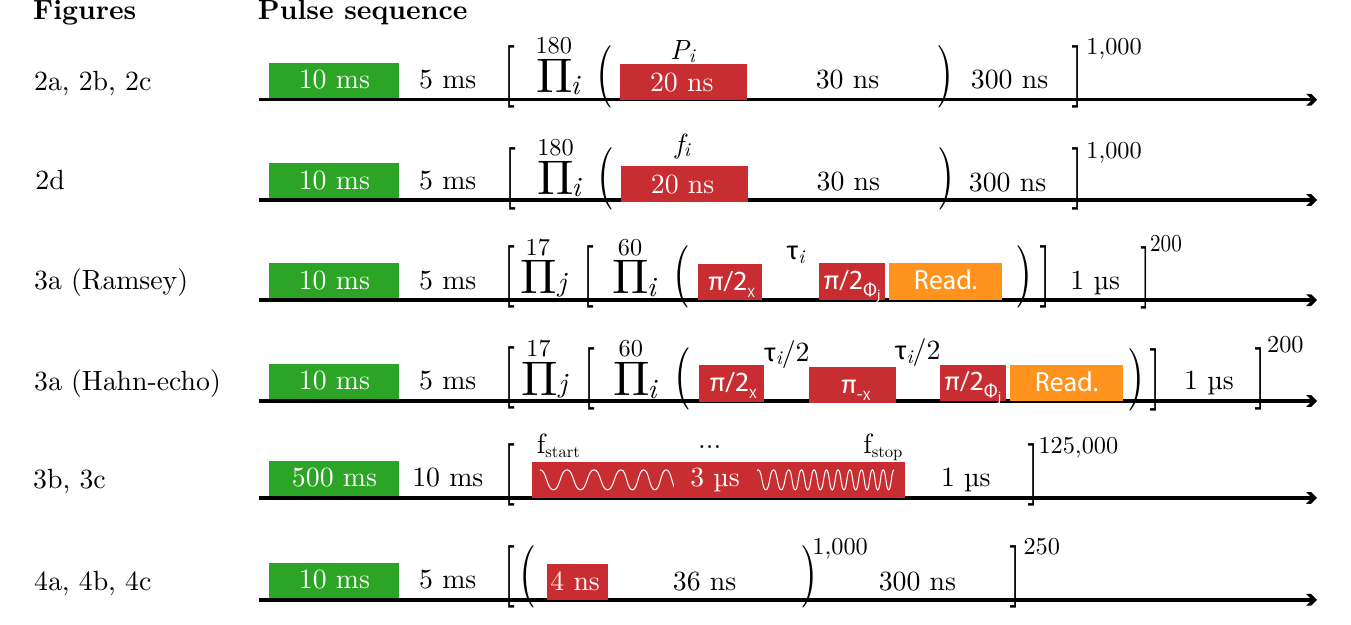}
    \caption{
        List of experimental control sequences employed for the main text figures. Green and red rectangles denote 532-nm and 619-nm excitation respectively. $P$ denotes the power of the beam, $f$ its frequency and $\tau$ periods of free precession of variable time length. Subscripts are used to indicate varying values of one of these variables in a sequence of pulses. Pulses are labeled with a subscript indicating their rotation axis.
    }
    \label{fig:pulse_sequence}
\end{figure*}

Figure \ref{fig:pulse_sequence} shows the pulse sequences employed for all measurements in the main text. Green rectangles denote the 532-nm off-resonant excitation used to reset the emitter into the bright state. The typical duty cycle for off-resonant excitation is approximately 50\%, with a continuous wave power of  approximately $10 \,\upmu$W. We find that much shorter sections of off-resonant excitation also serve to reset the emitter, but choose a longer duration, as this interval of the sequence is also used to lock the EOM (see section \ref{sec:Setup}). 
The 300 ns to 1 $\upmu$s sections at the end of each iteration are included to trigger the arbitrary waveform generator.

In Figs 2 and 3 of the main text, where we vary the detuning and phase of the 619-nm resonant excitation, we blue-detune the laser by 2 GHz from the optical transition. This 2 GHz detuning is bridged with an EOM, which provides sinusoidal amplitude modulation of the laser with variable amplitude, frequency and phase. The resulting laser field acquires two first-order sidebands where the lower sideband is resonant or only slightly detuned from the optical transition of the emitter, while the upper sideband is $\sim4$ GHz detuned and thus has negligible effect. In Fig. 4, we lock the laser frequency to be on resonance with the transition and drive the EOM with square pulses without sinusoidal modulation. 

During the readout sections (Read.) the emitter spontaneously decays and its fluorescence is collected. The length of the readout section is varied to compensate for the changing free precession time $\tau$ to keep each shot of the Ramsey and Hahn-echo experiments at a constant length of 50 ns. Within the readout section we then select a region of constant length for which we integrate the fluorescence counts.

\section{Master equation modelling}

We now describe a master equation model of a two-level system, that can be fit to the Rabi measurements presented in Fig~2a, b \& c and Ramsey \& Hahn-echo contrast decays in Fig~3a. In this model, the von Neumann equation acquires a non-unitary term, known as the Lindbladian super-operator, which models Markovian decoherence dynamics \cite{Lindblad1976}

\begin{equation}
\frac{\partial \rho}{\partial t} = \frac{-i}{\hbar} [H,\rho] + \sum_i c_i\rho c_i^\dagger -\frac{1}{2}(c_i^\dagger c_i\rho + \rho c_i^\dagger c_i),
\end{equation}
where $\{c_i\}$ are the set of collapse operators and $\rho$ is the density operator. The unitary evolution is given by the Hamiltonian

\begin{equation}
H=\frac{\Omega}{2}\sigma_x + \frac{\delta}{2}\sigma_z,
\end{equation}
where $\sigma_i$ is the $i$'th Pauli matrix, $\Omega$ the Rabi rate, $\delta$ the detuning and $\ket{0}$ ($\ket{1}$) is the ground-state (excited-state). The collapse operators are given by
\begin{align}
c_1&=\sqrt{\Gamma_{0}} \ket{0}\bra{1},\\
c_2&=\sqrt{\frac{\Gamma_\text{PD}}{2}}\sigma_z,
\end{align}
where $c_1$ describes spontaneous emission with rate $\Gamma_{0} = 1/T_1$ (where $T_1$ is the spontaneous emission lifetime), and $c_2$ describes pure dephasing with the rate $\Gamma_\text{PD}$, where $\Gamma_\text{PD}$ is the sum of the intrinsic pure dephasing rate $\Gamma_\text{PD}^\text{intrinsic}$ and the laser-induced dephasing rate $\Gamma_\text{PD}^\text{laser}$. To account for spectral diffusion, leading to inhomogeneous dephasing $T_2^*$, a phenomenological model is adopted. In this model, detunings are sampled from a normal distribution with standard deviation $\sqrt{2}/T_2^*$ where $T_2^*$ is a free fit-parameter for the Ramsey decay data presented in Fig. 3(a) of the main text and corresponds to the inhomogeneous dephasing lifetime \cite{barry2020sensitivity, appel_2021}. The different qubit evolutions are then averaged together, in the same way the spectral diffusion of the transition manifests itself in a time-averaged experiment. 
This quasi-static modelling of spectral diffusion is justified by the fact, that the time scale of spectral diffusion presented in Fig. 3(b) of the main text is much slower than the time scale of a single pulse sequence. Indeed, the model successfully recreates the observed dynamics in the Rabi oscillations of Fig. 2(a), 2(b) and 2(d) as well as the contrast decays in Ramsey and Hahn-echo spectroscopy in Fig. 3(a).

The fits in Fig~2(a), (b) \& (c) and in Fig~3(a) are all carried out using the same model. For Fig~2(a) a selection of Rabi oscillations are shown, with each curve corresponding to a different laser driving power. Here $\Gamma_0 = 1/T_1$ is fixed using the independently measured lifetime $T_1 = 7.44(20)$, $T_2^* = 4.54(2)$ ns is fixed from the fits to the Ramsey interferometry in Fig~3(a), and $\Gamma_\text{PD},\Omega$ are free fit parameters. The values of $\Omega$ obtained by fitting to all of the Rabi response curves are presented in Fig~2(b) and the Rabi quality factor is presented in Fig~2(c) where $Q=\Gamma_{\text{decay}} \Omega$, where $\Gamma_{\text{decay}} = \Gamma_0/2 + \Gamma_\text{PD}$ is the overall Rabi envelope decay rate. Thus, Fig~2(b) and (c) present all of the free fit parameters used in this model.

Fig~3(a) is fit using the same model as described above. However, we here decompose the pure dephasing rate into a laser-induced contribution $\Gamma_\text{PD}^\text{laser}$ and an intrinsic dephasing $\Gamma_\text{PD}^\text{intrinsic}$ owing to the emitter's inhomogenous linewidth broadening. During the laser pulses in Ramsey and Hanh-Echo spectroscopy, the qubit evolves with Rabi rate $\Omega$, spontaneously decays with rate $\Gamma_0$, and experiences pure dephasing with rate $\Gamma_\text{PD}=\Gamma_\text{PD}^\text{intrinsic}+\Gamma_\text{PD}^\text{laser}$. During the periods of free precession however, $\Omega=\Gamma_\text{PD}^\text{laser}=0$ due to the absence of a driving laser. Additionally, the described evolution is averaged over the inhomogeneous broadening quantified by $T_2^*$, which is a free fit parameter in the Ramsey experiment.

Ultimately, the free-fit parameters used to recreate the Ramsey contrast decay envelope in Fig~3(a) are $\Gamma_\text{PD}^\text{intrinsic} = 2\pi \times 6.99(75)$ MHz, $\Gamma_\text{PD}^\text{laser} = 2\pi \times 14.8(9)$ MHz and $T_2^* = 4.54(2)$ ns.
The free-fit parameters used to recreate the Hahn-echo contrast decay envelope in Fig~3(a) $\Gamma_\text{PD}^\text{intrinsic} = 2\pi \times 6.39(14)$ MHz and $\Gamma_\text{PD}^\text{laser} = 16.0(3)$ MHz. 

\section{Measurements of background}

\begin{figure*}[h]
    \centering
    \includegraphics[width=0.6\textwidth]{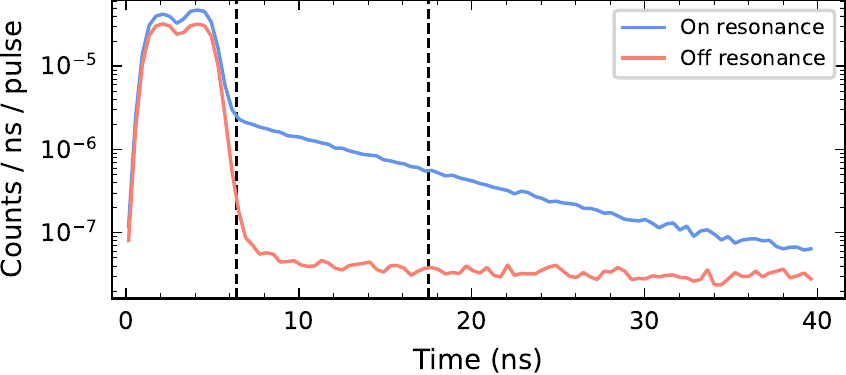}
    \caption{
        Time resolved ZPL fluorescence under excitation with $619-\text{nm}$ 4-ns-long pulses both on-resonance (blue) and off-resonance (red), at a repetition rate of 25 MHz. The region between the dashed lines corresponds to the fluorescence window collected in Fig. 4 of the main text.
    }
    \label{fig:decay}
\end{figure*}

To obtain a measurement of the signal-to-background ratio in our pulsed excitation experiment we excite the emitter on-resonance ($\delta = 0$) and off-resonance ($\delta = 6~\text{GHz}$) with a $4\text{-ns}$-long square pulse. Figure \ref{fig:decay} shows the collected ZPL fluorescence in both protocols, from which we extract a signal-to-background ratio of 23.91(3) in the region highlighted. Maximally extending the collection region to encompass the whole decay visible in Fig. \ref{fig:decay} reduces the signal-to-background ratio to 13.27(2).

We note that the small feature in the excitation pulse is a result of the generation protocol. As our microwave amplifiers cannot sustain a DC bias we first apply a $2$-$\text{ns}$ pulse of positive polarity  with an amplitude matching the half-wave voltage of the EOM, immediately followed by another pulse of equal length and amplitude but opposite polarity.

\section{Theoretical analysis of Hong-Ou-Mandel experiments}

\subsection{Effects of optical dephasing}

We now compute the expected measurable indistinguishability given our characterisation of the spectral diffusion through Ramsey and Hahn-echo spectroscopy. We follow the treatment of Kambs et al. closely \cite{kambs2018homanalysis} and reduce it to the specific case considered here: consecutive photons from a single emitter.

The emitter's radiated field $E_k(t)$ can be expressed in 2nd quantization by the creation operator for the optical mode weighted by, effectively, the amplitude of the probability distribution of the decay

\begin{equation}
E_k^+(t) = \zeta_k(t) a_k \quad \text{, and therefore} \quad E_k^-(t) = \zeta_k(t)^* a_k^\dagger\;,
\end{equation}
where $a_k^\dagger$ and $a_k$ are respectively the creation and annihilation operators for the photonic mode $k$. The weighting function $\zeta_k(t)$ is described for spontaneous decay with a lifetime $T_1$ and angular frequency $\omega$ as a one sided exponential function given by

\begin{equation}
\zeta_k(t) = \frac{1}{\sqrt{T_1}} H(t) \exp\left(-\frac{t}{2T_1}\right) \exp\left(- i \omega t\right) \exp\left(- 2 \pi i \Phi(t) \right)
\end{equation}
where $H(t)$ is the Heaviside function, defined to be 1 for positive values of its argument and otherwise zero, and the prefactor of $1/\sqrt{T_1}$ ensures wavefunction normalisation according to $\int_{-\infty}^{\infty} |\zeta_k(t)|^2 dt = 1$. The final term, $\exp\left(- 2 \pi i \Phi(t) \right)$ corresponds to pure dephasing, modelled as a fast time-varying phase.

A beamsplitter (BS) can now be considered as a unitary transformation between its two input modes and its two outputs modes, given explicitly by 

\begin{align}
U = \frac{1}{\sqrt{2}} \begin{bmatrix}
1 & 1 \\
1 & -1 
\end{bmatrix}
.
\end{align}

\begin{figure*}
  \centering
  \includegraphics[width=1.0\textwidth]{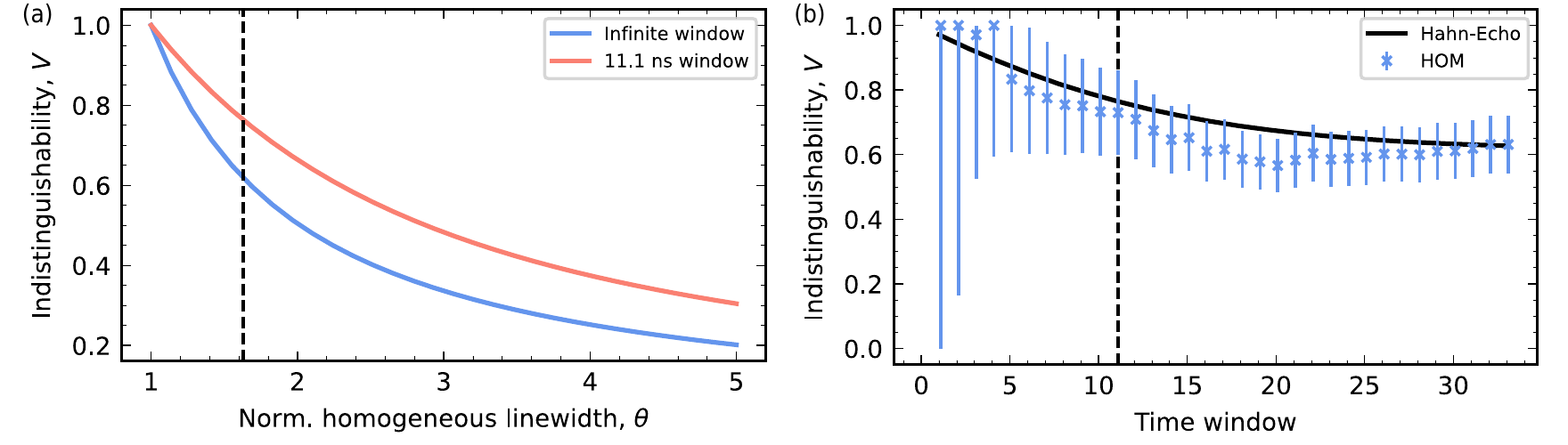}
  \caption{
      (a) Numerical simulation of the expected photon indistinguishability both for an infinite collection window (blue) and for the 11.1 ns window corresponding to 80\% of the fluorescence employed in this work (red). The dashed vertical line corresponds to the measured homogeneous linewidth through Hahn-Echo in the main text ($\theta$ = 1.63(4)). (b) Measured indistinguishability in Hong-Ou-Mandel as a function of the collection window (blue). The solid black line corresponds to a numerical simulation from the dephasing measured in Hahn-Echo. The dashed vertical line corresponds to the collection window highlighted in Fig. 4(a) of the main text.
  }
  \label{fig:hom_simulation}
\end{figure*}

Thus, the probability of measuring photons in separate output modes at times $t_0$ and $t_1$ can be mapped to $\zeta_1$ and $\zeta_2$ \cite{kambs2018homanalysis} according to

\begin{align}
P_{\text{joint}} &= |U_{2,1} U_{1,2} \zeta_1(t_1) \zeta_2(t_0)  + U_{2,2} U_{1,1} \zeta_2(t_1)  \zeta_1(t_0) |^2  \\
&= \frac{1}{4} |\zeta_1(t_1) \zeta_2(t_0)  - \zeta_2(t_1)  \zeta_1(t_0) |^2\;.
\end{align}

We consider the case of photons from a single emitter such that $\zeta_1$ and $\zeta_2$ share the same $T_1$. As the interferometric delay is much shorter than the timescales of spectral diffusion, as shown in the main text, there is also no difference in the central frequency of emission $\omega$. The source of indistinguishability is then the pure dephasing accounted by the time-varying phases $\Phi(t)_1$ and $\Phi(t)_2$. Thus \cite{kambs2018homanalysis}

\begin{equation}
P_{\text{joint}}(t_0, t_1) = \frac{1}{T_1^2} f(t_0, t_1) g(t_0, t_1)
\end{equation}
where

\begin{align}
f(t_0, t_1) &= \frac{1}{4} \exp\left(-(t_1 - t_0) / T_1\right) (2 + h(t_0, t_1)) \\
g(t_0,t_1) &= H(t_0) H(t_1) \exp(-2t_0/T_1) \\
h(t_0,t_1) &=  - 2 \cos(\Delta\Phi_1 - \Delta\Phi_2) 
\end{align}
and $\Delta\Phi_i = \Phi_i(t_1) - \Phi_i(t_0) $.

In an experimental measurement, the probability is averaged over many iterations. It can be shown that $h$ follows a correlation function characterised by a coherence time inversely proportional to the pure dephasing rate $\Gamma_\text{PD}$ \cite{kambs2018homanalysis}:

\begin{equation}
\expval{h(t_0,t_1)} = - 2 \exp(- 2 \Gamma_\text{PD} |t_1 - t_0|),
\end{equation}
where $\expval{}$ denotes an ensemble average. The probability of detecting photon coincidences is then given by

\begin{equation}
P_\text{coinc} = \int\limits_{0}^{t_\text{max}} \int\limits_{0}^{t_\text{max}}
dt_0 dt_1 \expval{P_\text{joint}(t_0,t_1)},
\end{equation}
where we have introduced a finite collection window extending to time $t_\text{max}$. The visibility can then be extracted by comparing the coincidences to those for distinguishable photons ($h(t_0,t_1)=0$). 

This model is employed in Fig S8(a) to calculate the dependence of the photon indistinguishability on the pure dephasing rate. As this only depends on the ratio of the pure dephasing rate to the transform-limited linewidth, the indistinguishability is presented as a function of the normalised homogeneous linewidth  $\theta = 1 + 2 \Gamma_\text{PD} / \Gamma_0$. In panel (b) the photonic indistinguishability is presented as a function of the length of the collection window ($t_\text{max}$). For an infinite collection window we compute, from our measured pure dephasing rate, a theoretical indistinguishability of 63.2(4)\%. This is in good agreement with our experimentally measured result of 63(9)\%.

\subsection{Effects of experimental imperfections}
We now consider how the measured Hong-Ou-Mandel visibility is reduced by experimental errors, namely imperfect beamsplitters (BS), imperfect interferometric visibility, and dark counts \cite{branczyk2017hombasictheory}. An imperfectly balanced, lossless BS can be modelled by writing the mapping between the input modes (a,b) and output modes (c,d) as
\begin{align}
    a^\dagger &\rightarrow \sqrt{T} c^\dagger + \sqrt{R} d^\dagger \\
    b^\dagger &\rightarrow \sqrt{R} c^\dagger - \sqrt{T} d^\dagger
\end{align}
where the $\dagger$ superscript denotes field creation operators, and $T$ and $R = 1 - T$ denote the BS transmission and reflection coeffecients, respectively. In the case where the recombination BS (second BS in main text Fig. 4c) is imbalanced, the ideal input state $\ket{1}_a \ket{1}_b = a^\dagger b^\dagger \ket{0}_a \ket{0}_b$) gets mapped to

\begin{align}
\ket{\Psi} &= (\sqrt{T_2} c^\dagger + \sqrt{R_2} d^\dagger)  (\sqrt{R_2} c^\dagger - \sqrt{T_2} d^\dagger) \ket{0}_c \ket{0}_d \\
&=  \sqrt{T_2 R_2} c^\dagger c^\dagger - T_2 c^\dagger d^\dagger + R_2 d^\dagger c^\dagger - \sqrt{R_2 T_2} d^\dagger d^\dagger \ket{0}_c \ket{0}_d \\
&=  \sqrt{2 T_2 R_2} \ket{2}_c \ket{0}_d - (T_2 - R_2) \ket{1}_c \ket{1}_d - \sqrt{2 T_2 R_2} \ket{0}_c \ket{2}_d,
\end{align}
where $T_2$ and $R_2$ are the transmission and reflection of the recombination BS, respectively. 
The probability of detecting a photon coincidence on opposite detectors when in the parallel polarisation configuration then becomes
\begin{align}
P_\parallel = |\bra{1}_c\bra{1}_d\ket{\Psi}|^2 = (T_2 - R_2)^2 = 1 - 4 R_2 T_2,
\end{align}
which is zero for the perfect case $R_2 = T_2 = 1/2$. 

Next, we consider the case of an imperfect interferometer due to imperfect spatial or polarisation overlap. Given a classical interference visibility $1-\epsilon$, the coincidence probability becomes \cite{tomm2021homthesis}

\begin{equation}
P_\parallel = 1 - 2 R_2 T_2 - 2 (1 - \epsilon)^2 V R_2 T_2.
\end{equation}

Finally, the introduction of a finite signal-to-background ratio $b$ affects the second order autocorrelation functions and leads to a measured $g^{(2)}(0) = 2/b$ in the limit $b\gg 1$. The effect from all three errors described above is contained in the expression \cite{tomm2021homthesis}
\begin{align}
P_\parallel=4 \left[ R_1 T_1 (1 - 2 R_2 T_2 - 2 (1 - \epsilon)^2 V R_2 T_2) 
+ 2 g^{(2)}(0) (1 - 2 R_1 T_1) R_2 T_2 \right],
\end{align}
where $V$ is the intrinsic indistinguishability of the emitter, and we additionally allow the first BS in the interferometer to have imbalanced reflection and transmission coeffecients given by $R_1$ and $T_1$, respectively. For the perpendicular case $P_\perp$, one can substitute a lack of indistinguishability ($V=0$) into the above expression. The raw measured visibility $V_\text{raw}$ is then given by $1 - P_\parallel/P_\perp$.

For the sake of simplicity, we evaluate the above expression for visibility in the limit of small deviations from a perfect experimental setup. Denoting the beamsplitter transmission and reflections as
\begin{equation}
T_i = \frac{1}{2} + \delta_i, \quad R_i = \frac{1}{2} - \delta_i,
\end{equation}
the intrinsic indistinguishability relates to $\text{V}_\text{raw}$ by \cite{tomm2021homthesis}
\begin{equation}
V = \frac{1}{(1-\epsilon)^2} \left( 1 + 2 g^{(2)}(0) \right) \left( 1 + 8 \delta_2^2 \right) V_\text{raw}.
\end{equation}
In our experiment the dominant correction terms are those involving $g^{(2)}(0) = 0.0836(4)$ and $\epsilon = 4(2)\%$. Here, the value of $g^{(2)}(0)$ is calculated from the signal-to-background ratio of 23.91(3) quoted in the main text, as this is a more precise estimate than directly measuring $g^{(2)}(0)$. The correction from the non-ideal recombination beamsplitter ($\delta_2 = 4(1)\%$) is much smaller.